\documentclass{article}

\PassOptionsToPackage{numbers,sort&compress}{natbib}
\usepackage[dblblindworkshop,final]{neurips_2026}
\workshoptitle{VLM4RWD}

\usepackage[utf8]{inputenc}
\usepackage[T1]{fontenc}
\usepackage{hyperref}
\usepackage{url}
\usepackage{booktabs}
\usepackage{amsmath}
\usepackage{amsfonts}
\usepackage{graphicx}
\usepackage{microtype}
\usepackage{xcolor}
\usepackage{multirow}
\usepackage{adjustbox}
\usepackage{float}
\usepackage{wrapfig}
\usepackage{fontawesome5}

\title{DrawVideo: Grounded and Faithful Multi-Shot Video Generation from Storyboard Keyframe Sketches}

\author{%
  Chuanzhi Xu\textsuperscript{1,*\textdagger}\quad
  Huiqi Liang\textsuperscript{1,*}\quad
  Bang Shi\textsuperscript{1}\quad
  Huiming Zhang\textsuperscript{1}\quad
  Guangcheng Lin\textsuperscript{1}\\
  \textbf{Yifan Xiao}\textsuperscript{1}\quad
  \textbf{Haodong Chen}\textsuperscript{1}\quad
  \textbf{Qiang Qu}\textsuperscript{1}\quad
  \textbf{Zhicheng Lu}\textsuperscript{2}\quad
  \textbf{Weidong Cai}\textsuperscript{1}\\[4pt]
  \textsuperscript{1}The University of Sydney\quad
  \textsuperscript{2}Charles Sturt University\\
  \textsuperscript{*}Equal contribution.\quad
  \textsuperscript{\textdagger}Corresponding author.\\[3pt]
  \texttt{chuanzhi.xu@sydney.edu.au}\quad
  \href{https://github.com/LouckXu/DrawVideo}{\faGithub\ Project Page}
}

\begin{document}

\maketitle

\begin{abstract}
Long video generation requires high-fidelity visual synthesis, coherent narrative organization, shot-level structure, and explicit user control. Existing text-to-video methods typically generate videos from a single long-form prompt, making it difficult for creators to directly control character pose, camera composition, spatial layout, and local motion. We propose DrawVideo, a sketch-guided and storyboard-driven framework that grounds multi-shot video generation in creator-provided spatial structure, appearance, and motion. Each shot is specified by a black-and-white sketch, an appearance prompt, and a motion prompt. DrawVideo first generates a structure-aligned reference keyframe, expands the motion description into derivative keyframes representing ordered action states, and then synthesizes local video clips between adjacent keyframes. We further introduce SketchLongVideo, to the best of our knowledge the first evaluation dataset of ordered multi-shot storyboards with aligned sketch, appearance, and motion conditions. Extensive experiments demonstrate strong structural controllability, appearance consistency, intra-shot visual stability, and motion alignment, providing an effective solution for director-oriented video creation.
\end{abstract}

\begin{figure}[H]
  \centering
  \includegraphics[width=\linewidth]{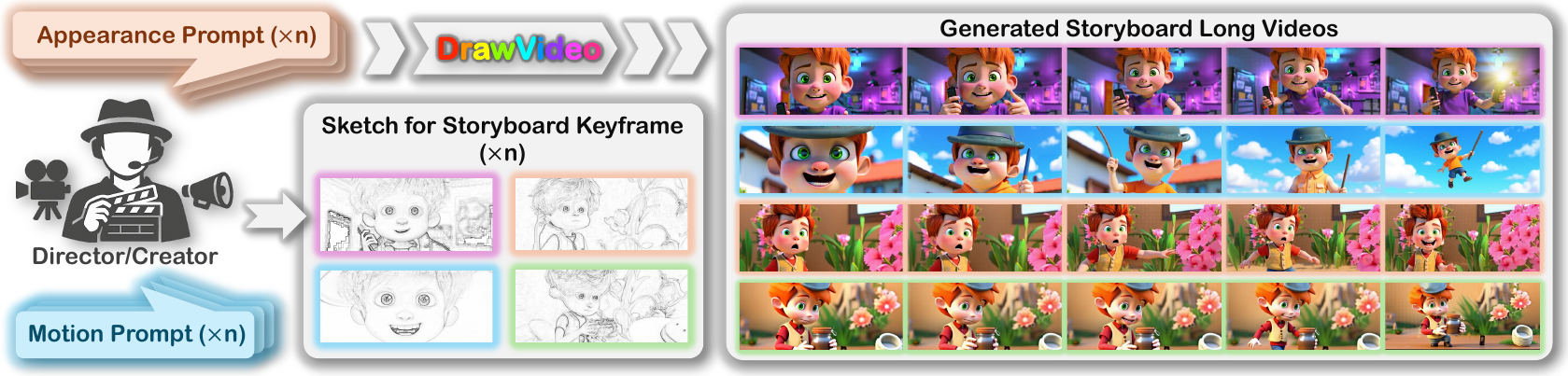}
  \caption{\textbf{Overview of DrawVideo.} A director provides a sketch keyframe and a pair of prompts describing appearance and motion for each storyboard shot. DrawVideo generates controllable and coherent storyboard shots and concatenates them into a long video.}
  \label{fig:teaser}
\end{figure}

\section{Introduction}
\label{sec:intro}

Imagine directors or video creators sketching storyboards, adding brief narrative descriptions, and automatically generating long videos aligned with their intended composition, semantics, and narrative. This would turn costly shot-by-shot production into a more efficient and controllable creative process.

Recent advances in diffusion-based and large-scale pretrained generative models have greatly improved image and video generation, enabling high visual fidelity, strong semantic alignment, and partial temporal consistency~\cite{VG0,VG1,VG2,VG3,VG4}. However, extending short-video generation to long videos with coherent narratives, cross-shot continuity, and strong controllability remains challenging. Existing methods mainly rely on hierarchical or segmented modeling, autoregressive extension, and cross-segment information propagation to improve scalability and long-range consistency~\cite{L4,L6,L2,L9,L7,L8,L1}. Nevertheless, their control signals are still largely limited to text prompts or implicit conditions, making it difficult to directly control shot composition, character layout, and shot-level structure for director-oriented creation.

Compared with generating an entire long video directly, a storyboard as an intermediate representation offers a more structured path that better matches real-world video creation workflows~\cite{ST3}. Prior studies have shown that storyboards play an important role in organizing story development, modeling cinematic language, and improving cross-shot consistency~\cite{ST4,ST1,ST5}. However, most existing methods still depend on text expansion, automatic shot planning, or data-driven scheduling, and provide limited explicit control over shot composition, character layout, and spatial structure. As a result, they still struggle to balance local controllability and intra-shot visual consistency.

At the same time, sketches, as a low-cost but highly expressive creative medium, have shown unique value in controllable visual generation. Compared with text, sketches can provide more direct geometric constraints on human pose, scene layout, and shot composition~\cite{SK7,SK3,SK1}. However, most existing sketch-driven methods focus on single-shot, short-duration, or local editing tasks and do not directly extend to storyboard-oriented long-video generation. Moreover, applying strong sketch constraints at every moment or across multiple keyframes often leads to identity drift, background changes, and stylistic instability~\cite{SK1,SK2,LongAnimation}. LVCD~\cite{LVCD} and LongAnimation~\cite{LongAnimation} rely on dense line-art sequences, in which line drawings are provided for the full frame sequence, and color these predefined sequences using a reference image. In contrast, director-oriented creation starts from sparse storyboard keyframes: creators specify shot-level composition, motion semantics, and narrative progression, while the generation system produces the intermediate motion. Existing line-art colorization methods therefore do not directly address storyboard-driven long-video generation.

Based on these observations, we propose \textbf{DrawVideo}, a sketch-guided long-video generation framework designed for director-oriented storyboard creation, as shown in Figure~\ref{fig:teaser}. DrawVideo allows users to create long storyboard sequences, where each shot is specified by three lightweight yet information-dense conditions: a black-and-white hand-drawn sketch, a motion prompt describing the local story, and an appearance prompt describing the characters and scene. These complementary visual and language conditions ground each generated shot in the supplied storyboard. DrawVideo processes them through Structured Prompt Decomposition, Sketch Coloring, Derivative Keyframes Generation, and First-last-frame Video Generation.
Our main contributions can be summarized as:
\begin{enumerate}
  \item We propose \textbf{DrawVideo}, a storyboard-driven long-video generation framework controlled by sparse sketches and paired appearance and motion prompts, providing explicit shot-level control over spatial structure, visual appearance, and local motion.
  \item We construct \textbf{SketchLongVideo}, to the best of our knowledge the first evaluation dataset of ordered multi-shot storyboards with aligned sketch, appearance, and motion conditions, following real-world workflows for shot-level planning and revision.
  \item Extensive experiments demonstrate that DrawVideo achieves strong structural controllability, appearance consistency, motion alignment, intra-shot visual stability, and overall long-video generation quality, validating the effectiveness of multimodal storyboard control.
\end{enumerate}

\section{Related work}
\label{sec:related}

\noindent\textbf{Long Video Generation.}
Existing long-video generation methods mainly follow two directions. One line of work reduces the difficulty of long-horizon modeling through hierarchical designs or segmented generation, such as coarse-to-fine generation and multi-segment temporal denoising~\cite{L4,L6}. The other line emphasizes autoregressive extension with memory mechanisms or progressive chunk-wise generation to improve temporal consistency over long durations~\cite{L2}. Recent studies have further advanced long-video generation from the perspectives of subject consistency, narrative planning, and multimodal control~\cite{L9,L7,L8,videodirectorgpt,seine}. Despite these advances, most existing methods still rely primarily on text prompts or latent conditions and therefore provide limited direct control over shot composition, character layout, and shot-level structure in director-oriented video creation.

\medskip
\noindent\textbf{Storyboard-based Video Generation.}
Compared with generating a full long video directly, a storyboard provides a more structured intermediate representation that better matches real-world creative workflows~\cite{ST3}. Prior work has explored storyboard understanding and generation through dataset construction, cross-modal retrieval, shot attribute modeling, and multi-shot narrative generation~\cite{ST4,ST1,ST5}. However, most existing approaches still rely on text expansion, automatic shot planning, or data-driven scheduling, but cannot achieve fine-grained and explicit control that faithfully reflects the creator's intended shot structure.

\medskip
\noindent\textbf{Sketch-based Video Generation.}
Sketches offer a low-cost and expressive interface for controllable visual generation, providing more direct geometric constraints than text on pose, layout, and composition. Sketch-to-image generation from abstract hand-drawn inputs shows the effectiveness of sketches as structural guidance~\cite{SK7}, while sequential sketch generation and human-model co-drawing reveal the procedural and interactive nature of sketching~\cite{SK10,SK5}. At the video level, existing methods study sparse-sketch-driven portrait or human video generation~\cite{SK3,SK6}, keyframe-based sketch-guided video generation and editing~\cite{SK1}, static-sketch-to-animation generation~\cite{SK12,SK2}, sketch video synthesis, spatiotemporal sketch representation, and sketch-guided inbetweening~\cite{SK4,SK8,SK9}. Existing methods remain focused on single-shot, short-duration, or local tasks, while multi-shot resources use reference images, captions, or frame pairs rather than aligned sketch, appearance, and motion conditions~\cite{AnimeShooter,ST5}.

\section{Construction of SketchLongVideo Dataset}
\label{sec:dataset}

\subsection{Motivation \& Overview}

\begin{figure}[!t]
  \centering
  \includegraphics[width=0.80\linewidth,pagebox=cropbox]{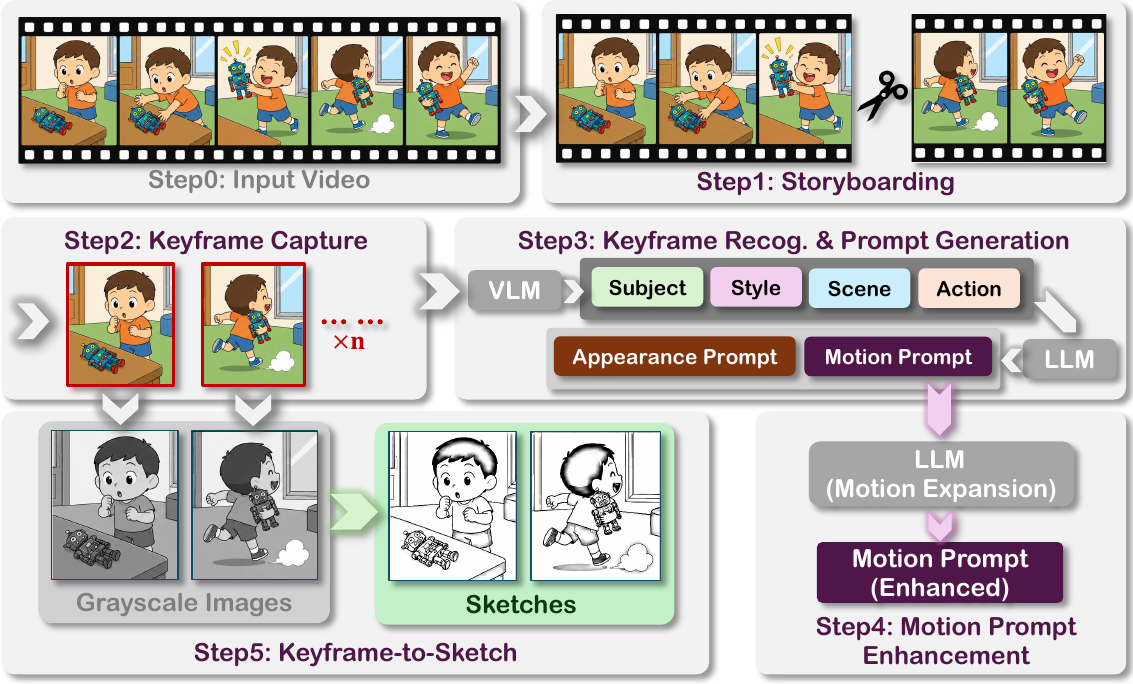}
  \caption{SketchLongVideo video-based dataset construction pipeline.}
  \label{fig:dataset_pipeline}
\end{figure}

DrawVideo targets sketch-guided storyboard-driven long-video generation. Instead of generating an entire long video from a single long-form text prompt, it follows the real-world directing workflow by decomposing a video into independently controllable shot-level units, i.e., storyboards. Each shot is represented by three conditions: a black-and-white sketch, an appearance prompt, and a motion prompt. The sketch controls pose, composition, and spatial layout. The appearance prompt specifies character identity, scene content, and visual style. The motion prompt describes shot-level action semantics. To support this setting, we introduce SketchLongVideo, to the best of our knowledge the first evaluation dataset of ordered multi-shot storyboards with aligned sketch, appearance, and motion conditions. It follows practical workflows by letting creators specify and revise composition, appearance, and motion for each shot. Figure~\ref{fig:dataset_pipeline} gives an overview of the construction pipeline.

SketchLongVideo is constructed from three complementary data sources: publicly accessible online animation videos, animation videos derived from AnimeShooter~\cite{AnimeShooter}, and text-prompt-driven AI-generated keyframes. For video-based samples, we detect storyboard shots, extract representative keyframes, and convert them into \((\text{sketch},\text{appearance},\text{motion})\) triplets. For AI-generated samples, generated images are converted to keyframe sketches, while their generation prompts are normalized into appearance and motion conditions. In this way, SketchLongVideo covers both animation footage and controlled synthetic keyframe sequences, simulating the sketch-and-text inputs used in the directing process.

\subsection{Data Collection and Processing}

For the online-animation subset, we curate videos with high visual quality, limited compression artifacts, clear character subjects, distinct shot boundaries, and observable motion changes. The AnimeShooter-derived subset follows the same processing protocol and extends the coverage of long animation content. For the generated subset, a detailed character prompt first produces a reference image, which then guides action-varying storyboard keyframes while preserving identity-related details. The original generation prompts are retained as appearance and motion annotations.

For video-derived samples, PySceneDetect divides each video into ordered shots, and FFmpeg extracts representative keyframes~\cite{Storyboard1,Storyboard2,Storyboard3,Storyboard4}. LLaVA-OneVision recognizes the subject, visual style, scene, and visible action~\cite{Storyboard5}. Qwen2.5 then organizes these results into an appearance prompt and a motion prompt~\cite{yang2024qwen25}. The motion prompt is further expanded into a local action narrative. Finally, a deterministic grayscale, inversion, erosion, and color-dodge process converts each color keyframe into a structural sketch. Full processing details and examples from all three sources are provided in Appendix~\ref{app:dataset_details}.

\subsection{Analysis of Data Statistics}

SketchLongVideo contains 1,233 aligned triplets forming 126 long-video generation sequences. The self-collected, AnimeShooter-derived, and AI-generated subsets contribute 201, 932, and 100 triplets, respectively, forming 20, 96, and 10 sequences. Most sequences contain around ten keyframes. The AnimeShooter-derived subset provides the largest portion of long animation content, while the AI-generated subset complements video-derived data with identity-consistent and controllable character sequences.

\section{Methodology -- DrawVideo Framework}
\label{sec:method}

\begin{figure}[t]
  \centering
  \includegraphics[width=\linewidth]{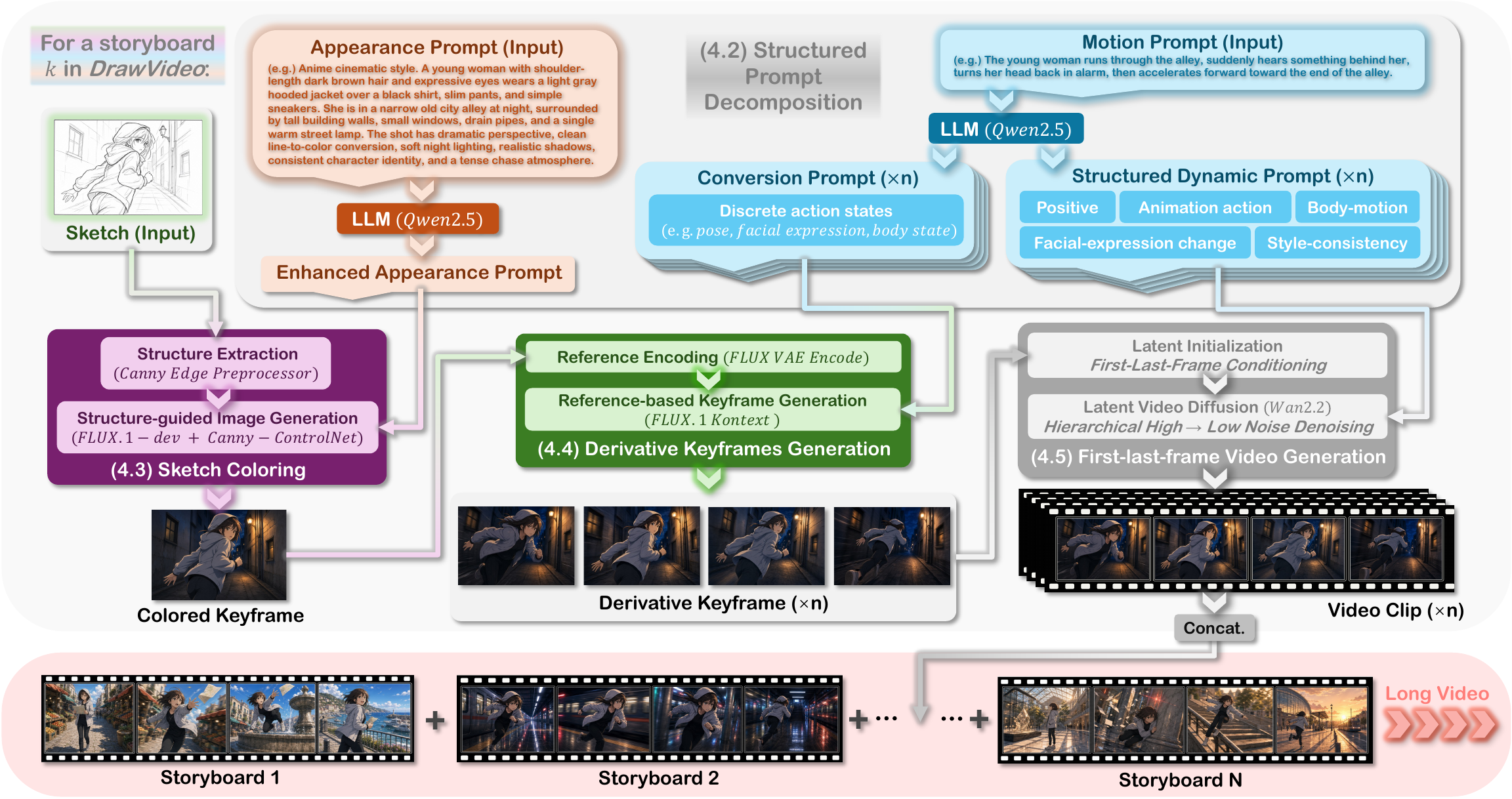}
  \caption{\textbf{Framework architecture of DrawVideo.} DrawVideo progressively generates controllable long videos from storyboard sketches and prompts.}
  \label{fig:method}
  \vspace{-6pt}
\end{figure}

\subsection{Overview}

DrawVideo is a storyboard-driven framework for controllable long-video generation from keyframe sketches and text prompts. It decomposes a long video into multiple independently controllable storyboard shots and adopts a hierarchical ``global multi-shot, local single-sketch'' generation strategy. Given a storyboard consisting of \(N\) shots, the input is \(\mathcal{B}=\{(S_k,A_k,M_k)\}_{k=1}^{N}\), where the \(k\)-th shot is specified by a sketch \(S_k\), an appearance prompt \(A_k\), and a motion prompt \(M_k\).

As shown in Figure~\ref{fig:method}, DrawVideo processes each shot through prompt decomposition, sketch coloring, derivative keyframe generation, and video synthesis. It first generates an initial keyframe from the input sketch and appearance prompt as the shot-level appearance anchor. The motion prompt is then converted into derivative keyframes representing discrete action states. Adjacent keyframes are finally used to generate short video clips, which are concatenated across shots to form the final long video. This design decouples structural control, appearance consistency, action-state control, and temporal motion synthesis, balancing director-level controllability with intra-shot visual consistency.

\subsection{Structured Prompt Decomposition}

The Qwen2.5-based Structured Prompt Decomposition module organizes semantic conditions for different generation stages. The appearance prompt is first standardized and enhanced with character identity, clothing, scene, camera composition, visual style, and consistency constraints. The enhanced prompt is used for sketch coloring, enabling the initial keyframe to preserve the sketch structure while producing a stable and stylistically consistent appearance.

The motion prompt is decomposed into exactly five ordered pairs, \(\mathcal{P}_{M}(M_k)=\{(C_k^i,D_k^i)\}_{i=1}^{5}\).
Conversion prompt \(C_k^i\) describes a discrete action state, such as a change in pose, facial expression, or body state, and is used to generate a derivative keyframe. Structured dynamic prompt \(D_k^i\) guides the transition from \(I_k^{i-1}\) to \(I_k^i\) through positive visual, animation action, body motion, facial-expression change, and style-consistency components. Each pair therefore has a one-to-one temporal correspondence with one derivative keyframe and one local video clip.

Because a still frame specifies an action state but cannot represent the intervening motion, pairing the two prompt types links each image target to the local transition that reaches it. Each storyboard's five states follow its own supplied motion. Appendix~\ref{app:prompt_details} provides the output schema and an illustrative five-state example.

\subsection{Sketch Coloring}

\begin{wrapfigure}{r}{0.50\textwidth}
  \centering
  \vspace{-8pt}
  \includegraphics[width=\linewidth]{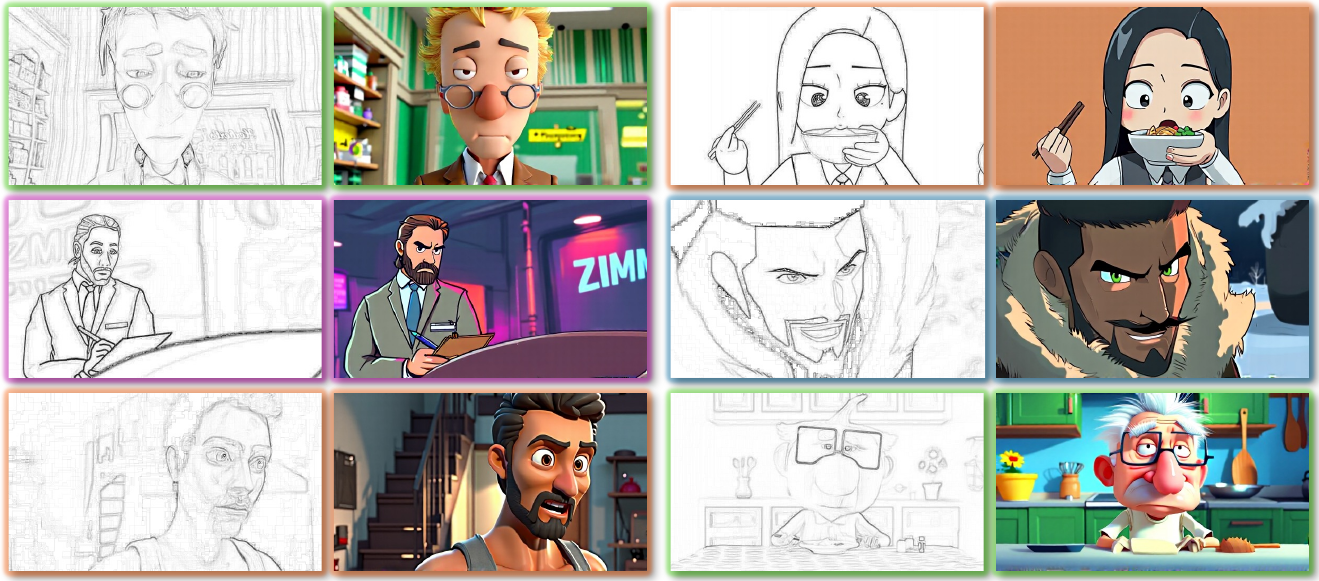}
  \caption{\textbf{Examples of sketch-to-keyframe coloring.} The colored reference keyframes retain the pose and composition of the input sketches.}
  \label{fig:sketch_coloring}
  \vspace{-8pt}
\end{wrapfigure}

The Sketch Coloring module converts each input storyboard sketch into a colored reference keyframe \(I_k^0=\mathcal{C}(S_k,\hat{A}_k)\). The sketch is first converted into a Canny edge map~\cite{canny1986computational}, which preserves the sparse contour structure without adding extra texture or shading cues. The edge map is then used as structural guidance for FLUX.1-dev~\cite{flux2024} through a Canny-ControlNet module based on ControlNet~\cite{zhang2023adding}, while the enhanced appearance prompt provides text conditioning.
The generated keyframe preserves the sketch-defined pose, composition, camera framing, and spatial layout while producing a fully colored and stylistically consistent image. It also serves as the appearance anchor for the shot, establishing character identity, clothing colors, scene layout, and visual style for the following stages. Examples are shown in Figure~\ref{fig:sketch_coloring}, and implementation settings are provided in Appendix~\ref{app:generation_details}.

\subsection{Derivative Keyframes Generation}

After obtaining the reference keyframe \(I_k^0\), DrawVideo generates five derivative keyframes representing major action states within the shot. The system uses \(I_k^0\) as the appearance reference and applies conversion prompts for reference-conditioned keyframe generation with FLUX.1 Kontext~\cite{labs2025flux1kontextflowmatching}, written as \(I_k^i=\mathcal{K}(I_k^0,C_k^i)\) for \(i=1,\ldots,5\).
The reference keyframe is encoded into the latent space through a VAE encoder and injected through ReferenceLatent conditioning. All five derivative keyframes are independently generated from this common reference. This keeps an appearance change or artifact in one derivative from being reused as the reference for later states. The reference supplies identity, scene, and style context, while each conversion prompt specifies a distinct target action. Temporal order is introduced when adjacent keyframes are paired for video synthesis. The detailed configuration is provided in Appendix~\ref{app:generation_details}.

\subsection{First-last-frame Video Generation}

For the \(k\)-th shot, DrawVideo constructs adjacent keyframe pairs \((I_k^{i-1},I_k^i)\) and combines each pair with its structured dynamic prompt \(D_k^i\). Wan2.2-I2V-A14B then generates a local video clip using first-last-frame conditioning~\cite{wan2025}, written as \(V_k^i=\mathcal{G}(I_k^{i-1},I_k^i,D_k^i)\) for \(i=1,\ldots,5\).
The high-noise stage generates coarse temporal dynamics and motion transitions, while the low-noise stage refines appearance consistency and visual details. Local synthesis assigns the video backbone a short transition with explicit visual start and end states. The derivative keyframes provide the action sequence before video synthesis, allowing the video model to focus on the intermediate motion between adjacent states. The paired dynamic prompt specifies how that local transition should unfold while retaining the shot's appearance conditions. The local clips are concatenated in temporal order to form each shot, and the shots are composed according to the storyboard order to obtain the final long video. The overall duration grows with the number of storyboard shots supplied by the creator, while shot boundaries follow the intended cuts in the storyboard. Full sampling and video settings are provided in Appendix~\ref{app:generation_details}.
Each storyboard shot remains an independent generation unit defined by its own \((S_k,A_k,M_k)\) triplet. A creator can revise the structure, appearance, or motion condition of one shot and regenerate it while retaining the remaining storyboard order. The pipeline operates entirely at inference time, and the colored reference keyframe and derivative action states are available before video synthesis. The revised shot is then reinserted into the same storyboard sequence.

\begin{table}[b]
\centering
\scriptsize
\setlength{\tabcolsep}{2.25pt}
\renewcommand{\arraystretch}{1.08}
\caption{Quantitative comparison by shot control, shot consistency, story alignment, local video quality, and auxiliary dynamic progression. Best results are shown in bold.}
\label{tab:results}
\begin{adjustbox}{width=\linewidth}
\begin{tabular}{lcccccccccc}
\toprule
Method &
\multicolumn{3}{c}{Shot Control} &
\multicolumn{2}{c}{Shot Consistency} &
\multicolumn{2}{c}{Story Alignment} &
\multicolumn{2}{c}{Local Video Quality} &
\multicolumn{1}{c}{Aux.} \\
\cmidrule(lr){2-4}\cmidrule(lr){5-6}\cmidrule(lr){7-8}\cmidrule(lr){9-10}\cmidrule(lr){11-11}
& LPIPS \(\downarrow\) & CLIP \(\uparrow\) & Edge-F1 \(\uparrow\)
& T-CLIP \(\uparrow\) & T-LPIPS \(\downarrow\)
& Static \(\uparrow\) & Story \(\uparrow\)
& Event \(\uparrow\) & Dynamic \(\uparrow\) & Progress. \(\uparrow\) \\
\midrule
DrawVideo
& \textbf{.4751} & .7806 & \textbf{.4664}
& \textbf{.9225} & .3830
& \textbf{.3781} & \textbf{.3111}
& \textbf{.7778} & \textbf{.6973} & .2493 \\
\midrule
SketchVideo (1KF)
& .6888 & .7640 & .3255 & .8360 & .5683 & .3285 & .2923 & .4444 & .5551 & .0304 \\
SketchVideo (2KF)
& .7204 & .7416 & .2649 & .8905 & .4626 & .2939 & .2764 & .5833 & .6114 & .2300 \\
VidSketch
& .6694 & .7051 & .2759 & .7430 & .5319 & .3672 & .2570 & .7037 & .6625 & .2721 \\
FlipSketch
& .5554 & .7103 & .1282 & .7508 & .7181 & .2239 & .2768 & .6667 & .6476 & .1628 \\
StoryDiffusion
& -- & -- & -- & .8314 & .5474 & .2859 & .2780 & .6667 & .6470 & \textbf{.4452} \\
\midrule
Wan2.2 (sketch+text)
& .4859 & \textbf{.7848} & .3541 & .8796 & .4893 & .3520 & .3023 & .6667 & .6542 & .1893 \\
Wan2.2 (text only)
& -- & -- & -- & .8788 & \textbf{.3651} & .2580 & .2441 & .3333 & .5070 & .2014 \\
\bottomrule
\end{tabular}
\end{adjustbox}
\end{table}

\section{Experiments \& Results}
\label{sec:experiments}

\subsection{Experimental Setup}
\label{sec:experimental_setup}

All experiments are conducted on the full SketchLongVideo dataset using a single NVIDIA RTX 5090 GPU for inference.

\noindent\textbf{Baselines.}
We compare against seven baselines. SketchVideo is evaluated in both single-keyframe (1KF) and two-keyframe (2KF) settings~\cite{SK1,yang2024cogvideox}. VidSketch receives one sketch, the full text condition, and the colorized reference frame required by its interface~\cite{SK2}. FlipSketch receives one sketch and the text condition~\cite{SK12}. StoryDiffusion is used as a text-only long-range generation baseline. Its story keyframes are animated by the same Wan2.2 backend~\cite{L9,podell2024sdxl,wan2025}. We further include Wan2.2 conditioned on sketch and text, and on text only. For all baselines that accept one text input, the appearance prompt and motion prompt are concatenated. Together, these baselines cover direct sketch-conditioned video generation, static-sketch animation, text-based long-range generation, and backbone-only generation without DrawVideo's intermediate representation. This grouping separates the effect of the input modalities from the effect of organizing them into reference and derivative keyframes. Additional implementation details are provided in Appendix~\ref{app:baseline_details}.

\noindent\textbf{Evaluation Protocol.}
We evaluate how generated videos align with the supplied storyboard conditions using ten metrics covering shot control, shot consistency, story alignment, local video quality, and observable dynamic progression. LPIPS, CLIP Image Similarity, and Edge-F1 evaluate the generated initial keyframe against the input sketch~\cite{zhang2018unreasonable,radford2021learning,canny1986computational}. Temporal CLIP and Temporal LPIPS measure semantic stability and perceptual drift within each shot. Static Alignment and Story Alignment compare generated visual content with the appearance and motion prompts. Event Completion and Dynamic Controllability evaluate whether local events are expressed and temporally controlled, following the event-level view of StoryEval and DEVIL~\cite{wang2024storyeval,liao2024devil}. The three shot-control measures are interpreted together: Edge-F1 directly evaluates contour agreement, while CLIP Image Similarity and LPIPS provide semantic and perceptual comparisons between the sketch and the completed reference keyframe. Dynamic Progression is reported separately as an auxiliary measure of visible temporal change rather than a standalone measure of motion correctness. The protocol follows the staged pipeline: shot control evaluates the spatial anchor created by Sketch Coloring, shot consistency evaluates whether this anchor is retained as motion develops, and story alignment and local video quality evaluate whether the appearance and motion conditions are expressed in the generated frames. Because the reference and derivative keyframes are explicit intermediate outputs, each group of metrics is connected to an observable generation stage. Detailed definitions are provided in Appendix~\ref{app:evaluation_details}.

\subsection{Main Comparisons}

Table~\ref{tab:results} shows that DrawVideo achieves the best performance on most metrics across shot control, shot consistency, story alignment, and local video quality. Compared with Wan2.2 (sketch+text), DrawVideo improves Edge-F1 from 0.3541 to 0.4664, Event Completion from 0.6667 to 0.7778, and Dynamic Controllability from 0.6542 to 0.6973. Temporal CLIP also increases from 0.8796 to 0.9225. These results jointly support contour-level faithfulness, intra-shot semantic stability, and the realization of motion-prompt events. They connect the spatial structure of the initial keyframe with the appearance and action conditions expressed across the generated shot.

The metrics also distinguish visible temporal change from condition-aligned motion. StoryDiffusion obtains a higher Dynamic Progression score than DrawVideo (0.4452 versus 0.2493), while DrawVideo achieves higher Event Completion (0.7778 versus 0.6667) and Dynamic Controllability (0.6973 versus 0.6470). Thus, the amount of visible change and the realization of the requested events capture complementary properties. Similarly, Wan2.2 (text only) obtains the lowest Temporal LPIPS (0.3651), whereas its Event Completion is 0.3333. Reading these measures together supports evaluating visual stability alongside whether the intended action is expressed. DrawVideo combines strong appearance and story alignment with event-level control, rather than optimizing any one temporal measure in isolation.

\subsection{Ablation Study}

The two Wan2.2 variants use the same video backbone as DrawVideo and compare the complete staged pipeline with direct conditioning. The text-only variant removes direct structural guidance, while the sketch-and-text variant retains the input modalities but omits appearance anchoring, action-keyframe expansion, and motion decomposition. These comparisons demonstrate the benefit of the complete staged pipeline over direct conditioning with the same video backbone. The reference keyframe establishes a colored visual state, while the ordered derivative keyframes turn the motion description into local transitions.

\subsection{Qualitative Evaluation}

\begin{figure}[t]
  \centering
  \includegraphics[width=0.97\linewidth]{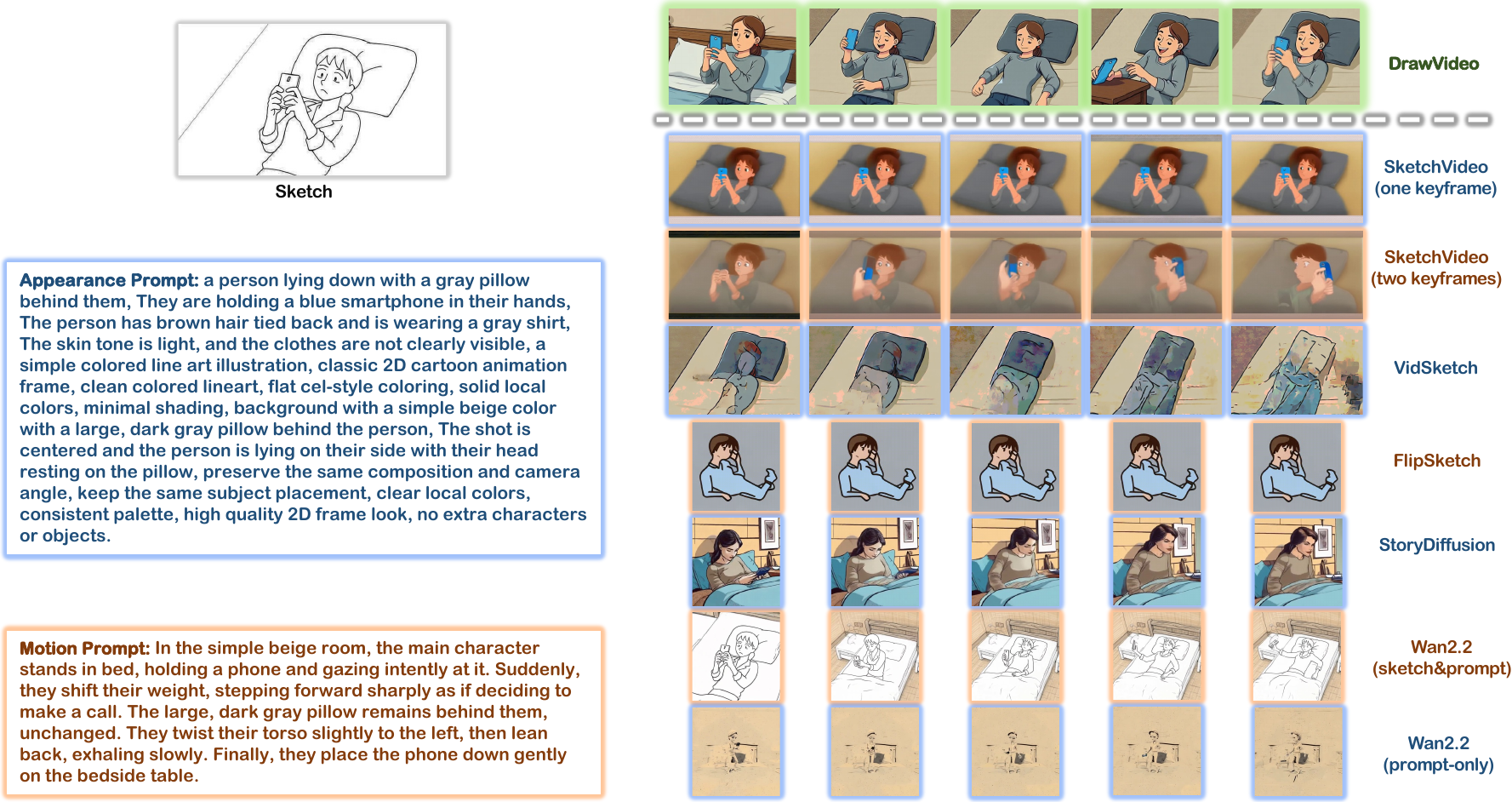}
  \caption{\textbf{Qualitative comparison on a phone-interaction storyboard.} Baseline settings follow Section~\ref{sec:experimental_setup}. One and two keyframes correspond to 1KF and 2KF in Table~\ref{tab:results}, and \emph{prompt} denotes the text condition.}
  \label{fig:comparison}
\end{figure}

\begin{figure}[t]
  \centering
  \includegraphics[width=0.97\linewidth]{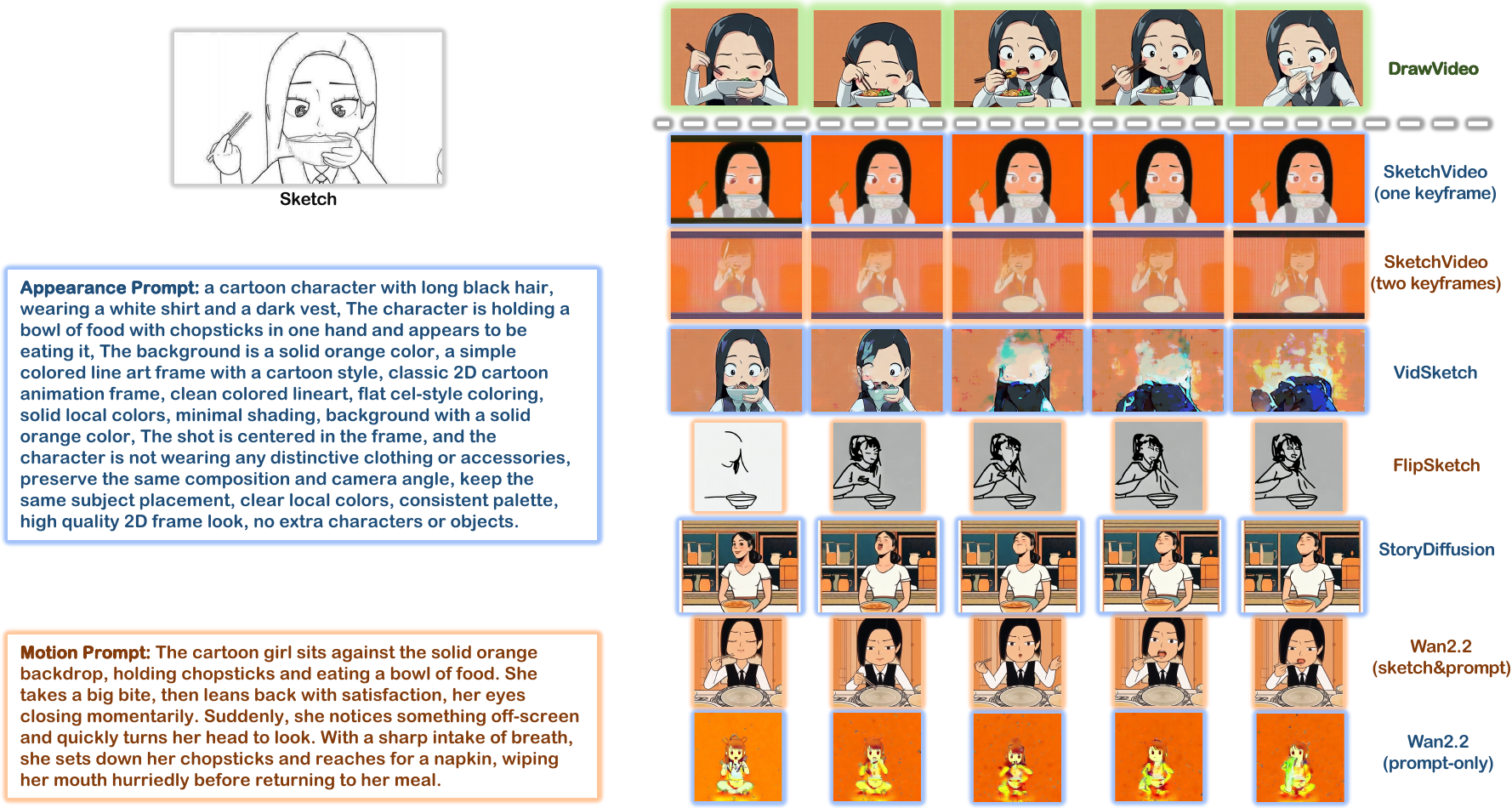}
  \caption{\textbf{Qualitative comparison on a fine-grained eating storyboard.} Baseline settings follow Section~\ref{sec:experimental_setup}. Method order and frame layout follow Figure~\ref{fig:comparison}.}
  \label{fig:comparison_eating}
\end{figure}

Figures~\ref{fig:comparison} and~\ref{fig:comparison_eating} compare DrawVideo with all baselines on storyboards featuring phone interactions and fine-grained character actions, respectively, using the settings in Section~\ref{sec:experimental_setup}. Both examples assess sketch-defined structure, appearance consistency, and motion-prompt alignment. The phone-interaction example shows changes in body pose and phone handling, while the eating example shows hand movements and facial-expression changes. Across both cases, DrawVideo's initial reference keyframes follow the supplied spatial structure, while the derivative keyframes carry ordered action states into the generated motion. The comparisons provide visual counterparts to the automatic metrics, relating the initial structure to appearance stability and action progression across the sequence.

\begin{figure}[t]
  \centering
  \includegraphics[width=0.97\linewidth]{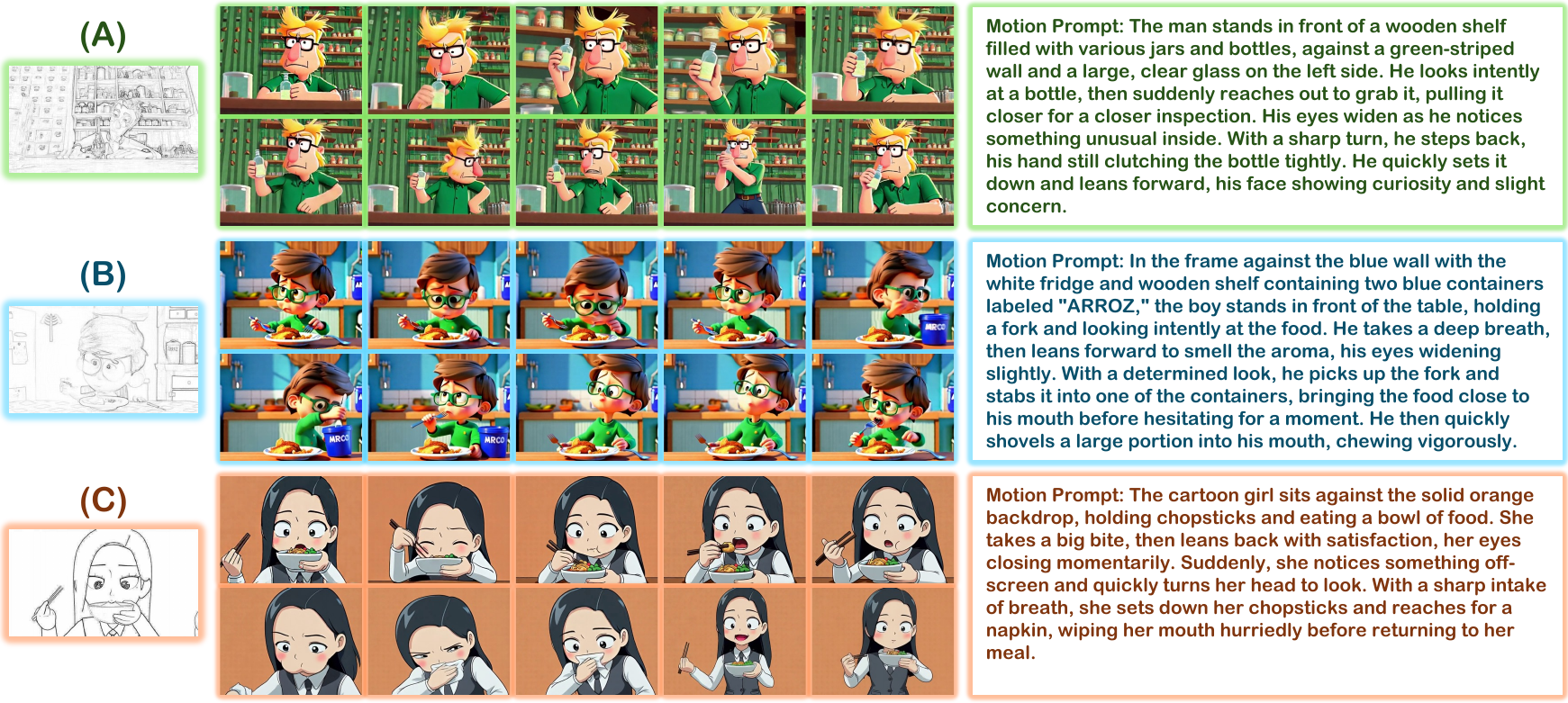}
  \caption{\textbf{Grounding across three generated shots.} The examples align sketch-specified structure and local motion prompts with sampled frames across different subjects, compositions, and actions.}
  \label{fig:consistency}
\end{figure}

As shown in Figure~\ref{fig:consistency}, DrawVideo preserves sketch-controlled structure while following the motion prompts across different subjects, compositions, and actions. The reference keyframes provide complete colored appearances, while the derivative keyframes carry visible action changes through the local clips. These examples also show how the three conditions are carried through the pipeline: sketch coloring establishes the spatial structure and complete visual appearance, all derivative keyframes share the same reference keyframe, and adjacent colored keyframes constrain each local transition. Together with Figures~\ref{fig:comparison} and~\ref{fig:comparison_eating}, this connects the quantitative comparison to visible behavior: the sketch controls spatial layout, while the shared reference and ordered motion conditions guide how each shot develops. Additional qualitative results are provided in Appendix~\ref{app:additional_results}.

Beyond final-video quality, DrawVideo exposes the colored reference and derivative keyframes as visible checkpoints before video synthesis. These intermediate states let creators inspect how the sketch, appearance prompt, and motion prompt are translated into a planned shot. Because each shot remains an independent generation unit, creators can revise its structure, appearance, or motion condition and regenerate that shot while preserving the remaining storyboard order.

\subsection{Human Evaluation Study}

We conduct a blind, randomized human evaluation. Participants view each storyboard's inputs and anonymized outputs in random order, rating structural faithfulness, appearance consistency, motion naturalness, controllability, and overall quality. DrawVideo receives the strongest overall evaluation in structural faithfulness, appearance consistency, storyboard controllability, and overall quality, while maintaining competitive motion naturalness. Together, the automatic metrics and human ratings assess both condition-level alignment and complete storyboard quality. The full protocol and scores are in Appendix~\ref{app:human_evaluation}.

\section{Conclusion}
\label{sec:conclusion}

We presented DrawVideo, a training-free framework for multi-shot video generation grounded in sketches and paired appearance and motion prompts. Its staged pipeline combines sketch-guided appearance completion, reference-conditioned action-state generation, and local video synthesis. SketchLongVideo provides ordered storyboards with aligned sketch, appearance, and motion conditions for evaluation. Experiments support the pipeline's effectiveness in structural faithfulness, intra-shot appearance consistency, and motion-prompt alignment. Colored reference and derivative keyframes expose the planned shot for inspection before synthesis, while independent shot generation supports local revision. Together, the framework and dataset provide a practical basis for controllable, storyboard-driven video creation.

\clearpage
% References

\clearpage
\appendix
\begin{center}
  {\LARGE\bfseries Appendix}
\end{center}
\vspace{0.5em}
\section{SketchLongVideo Details and Examples}
\label{app:dataset_details}

SketchLongVideo is built from three complementary sources: publicly accessible online animation videos, the AnimeShooter dataset~\cite{AnimeShooter}, and text-prompt-driven AI-generated keyframes. The two video-derived subsets follow the same pipeline, while the generated subset starts from identity-consistent character keyframes and retains the generation prompts as text conditions. Every sample uses a shared identifier to align one sketch, one appearance prompt, and one motion prompt, and sequence identifiers retain the storyboard order.

\noindent\textbf{Video-based preprocessing.}
PySceneDetect first divides each input video into shots using ContentDetector with a threshold of 25~\cite{Storyboard1,Storyboard3,Storyboard4}. FFmpeg extracts the center frame of each shot as its representative keyframe~\cite{Storyboard2}. Additional initial, transition, or ending frames are retained when a shot contains several visible action states. The extracted keyframes are resized to a width of 600 pixels while preserving their aspect ratios.

\medskip
\noindent\textbf{Prompt generation.}
For each video-derived keyframe, LLaVA-OneVision performs four role-specific queries for the subject, visual style, scene, and visible action~\cite{Storyboard5}. Qwen2.5 reorganizes these outputs into an appearance prompt containing subject, style, scene, and composition information and a motion prompt containing subject, scene, and action information~\cite{yang2024qwen25}. The initial motion prompt is expanded into a local action narrative while retaining the same subject, scene, and shot context. For the generated subset, the original generation prompts are normalized into the same two fields.

\medskip
\noindent\textbf{Sketch conversion and alignment.}
Each color keyframe is converted into a black-and-white structural sketch through grayscale conversion, intensity inversion, \(3\times3\) erosion, and color-dodge blending. Given grayscale image \(G\) and processed inverted image \(E\), the sketch is \(S(x)=\min\!\left(255,{G(x)\cdot255}/{(255-E(x)+\epsilon)}\right)\).
The final assembly retains samples only when the shared identifier matches across all three modalities and both text fields are non-empty. The resulting dataset contains 1,233 aligned triplets across 126 sequences and is used only for inference-time evaluation.

\begin{figure}[H]
  \centering
  \includegraphics[width=\linewidth]{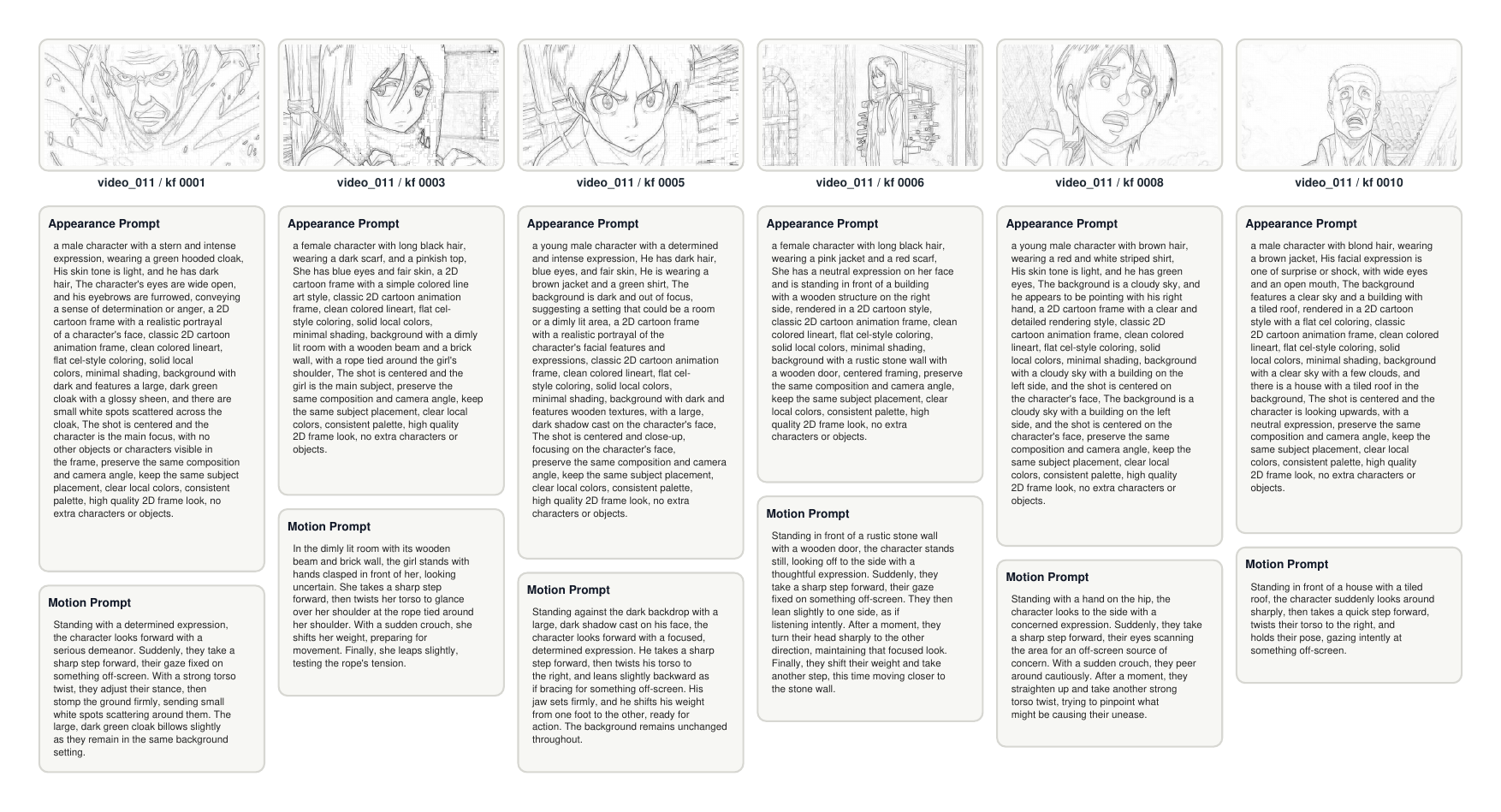}
  \caption{Examples from the self-collected online-animation subset. Each storyboard keyframe is represented by a structural sketch and paired appearance and motion prompts.}
  \label{fig:app_self_collected}
\end{figure}

\begin{figure}[H]
  \centering
  \includegraphics[width=\linewidth]{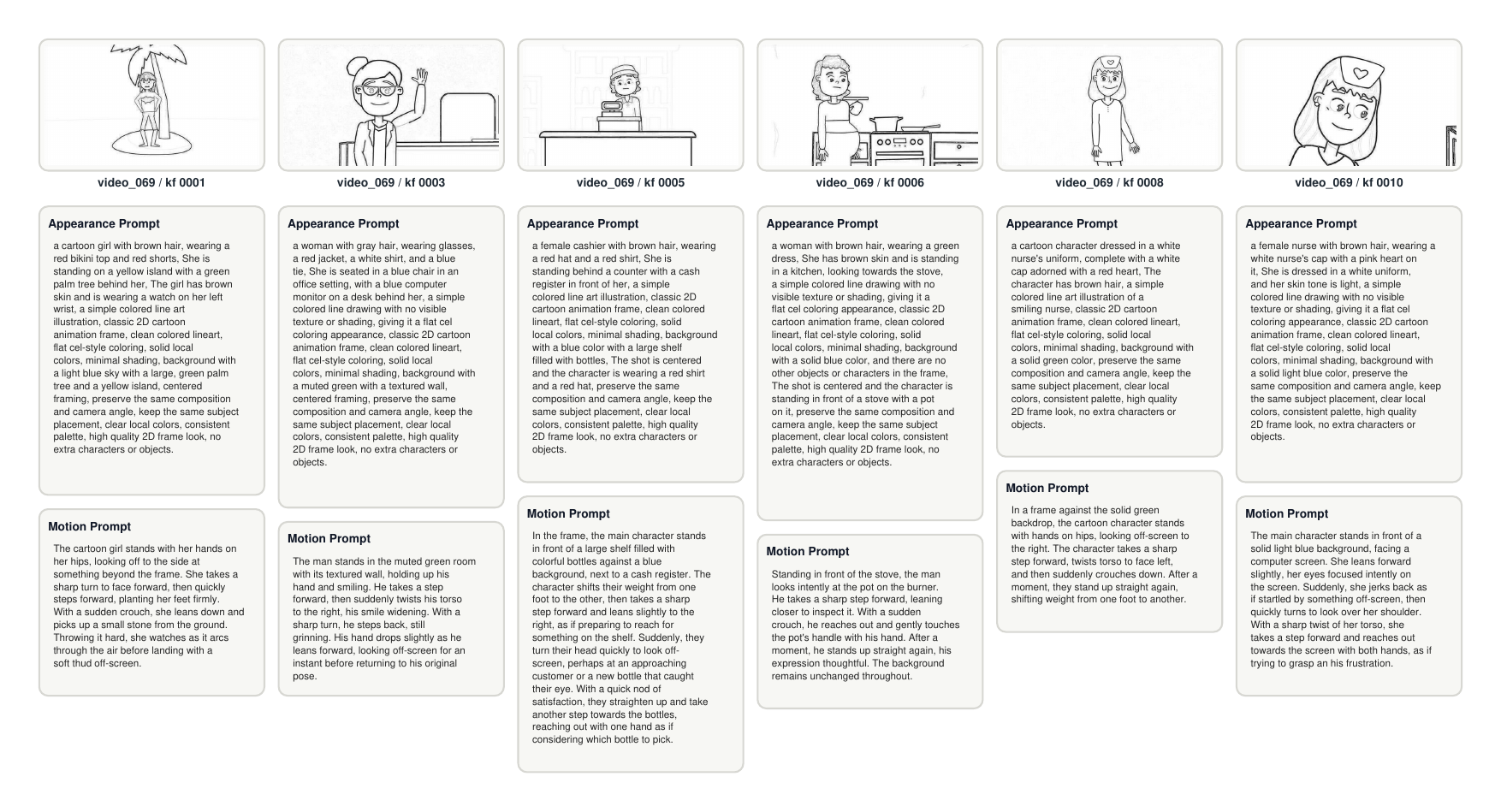}
  \caption{Examples from the AnimeShooter-derived subset, processed into the same aligned triplet format.}
  \label{fig:app_animeshooter}
\end{figure}

\begin{figure}[H]
  \centering
  \includegraphics[width=\linewidth]{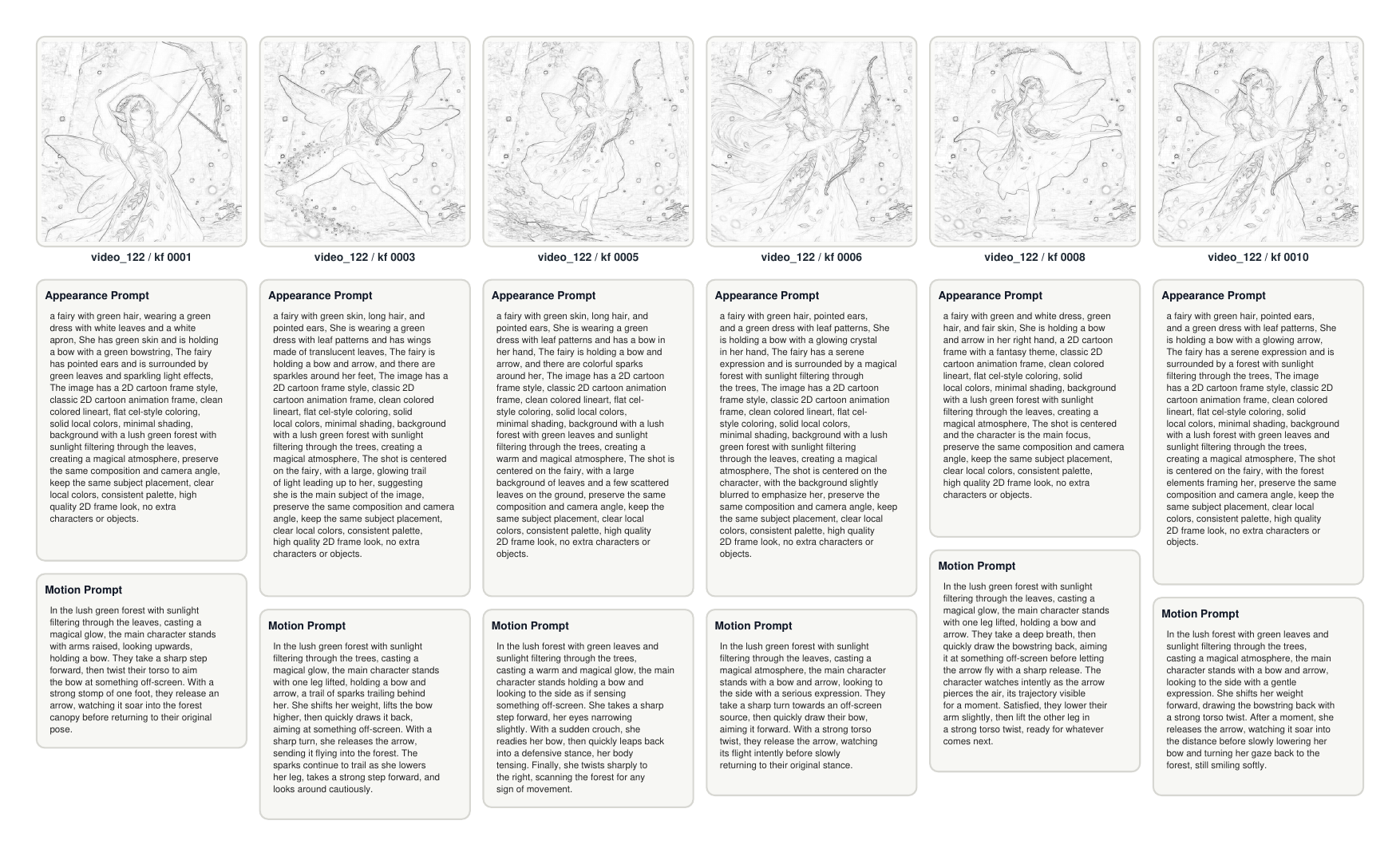}
  \caption{Examples from the AI-generated subset. Identity-consistent keyframes are converted into sketch conditions and paired with prompts derived from their generation descriptions.}
  \label{fig:app_ai_generated}
\end{figure}

Figures~\ref{fig:app_self_collected}, \ref{fig:app_animeshooter}, and \ref{fig:app_ai_generated} show examples from the three source subsets. Together, they illustrate the common sketch, appearance, and motion representation used throughout SketchLongVideo.

\section{Structured Prompt Decomposition Details}
\label{app:prompt_details}

Structured Prompt Decomposition transforms a coarse storyboard description into conditions for the following image and video stages. The appearance branch expands the appearance prompt with character identity, clothing, scene content, visual style, camera composition, and consistency terms. The motion branch produces five ordered pairs \(\{(C_k^i,D_k^i)\}_{i=1}^{5}\). Each conversion prompt \(C_k^i\) describes a discrete target action state, while its structured dynamic prompt \(D_k^i\) describes the transition ending at that state.

Each structured dynamic prompt contains five fields: a positive visual condition, animation action, facial-expression change, body motion, and style consistency. The output schema retains the original character name, scene, background, camera, and framing across all five stages. The generated prompts are stored in stage order and paired directly with the corresponding derivative keyframes and local clips.

\medskip
\noindent\textbf{Illustrative five-state example.}
An illustrative five-state decomposition of a hoop-dance motion is: looking upward while raising the hoop, stepping forward beside a cat, crouching, jumping with the hoop raised, and landing with a smile. These states specify the targets for the five conversion prompts. For the jump state, the conversion prompt describes the raised arms and airborne pose, while the paired dynamic prompt describes the transition from the preceding crouch through leg extension and upward body motion. The positive visual and style-consistency components retain the character, clothing, cat, scene, and rendering style. The action, body-motion, and facial-expression components specify the local changes.

\section{Generation Implementation Details}
\label{app:generation_details}

\noindent\textbf{Sketch Coloring.}
The sketch is converted to a Canny edge condition and injected into FLUX.1-dev~\cite{flux2024} through a FLUX-compatible Canny-ControlNet. The enhanced appearance prompt controls character identity, clothing, scene content, lighting, composition, and visual style. The settings for the results shown in Figure~\ref{fig:sketch_coloring} are summarized in Table~\ref{tab:app_sketch_config}.

\begin{table}[H]
\centering
\small
\caption{Implementation configuration of Sketch Coloring.}
\label{tab:app_sketch_config}
\begin{tabular}{ll}
\toprule
Component & Setting \\
\midrule
Image backbone & FLUX.1-dev \\
Text encoders & T5-XXL and CLIP-L \\
Structural preprocessor & Canny edge \\
Control module & FLUX Canny-ControlNet \\
Preprocessing resolution & 1024 \\
ControlNet strength & 0.95 \\
Control range & 0.0--1.0 \\
Sampler / scheduler & DPM++ 2M / SGM uniform \\
Sampling steps & 15 \\
CFG / FLUX guidance & 7.0 / 3.5 \\
Denoising strength & 0.8 \\
Output & Colored reference keyframe \(I_k^0\) \\
\bottomrule
\end{tabular}
\end{table}

\noindent\textbf{Derivative Keyframes Generation.}
FLUX.1 Kontext receives the reference keyframe through ReferenceLatent conditioning and one conversion prompt for each action state. All derivative keyframes are generated independently from \(I_k^0\); a derivative keyframe is never used as the appearance reference for the next one. This shared-reference design prevents recursive error accumulation and retains the same character, scene, and visual style across different action states. Table~\ref{tab:app_derivative_config} gives the configuration.

\begin{table}[H]
\centering
\small
\caption{Implementation configuration of Derivative Keyframes Generation.}
\label{tab:app_derivative_config}
\begin{tabular}{ll}
\toprule
Component & Setting \\
\midrule
Generation backbone & FLUX.1 Kontext \\
Conditioning mechanism & ReferenceLatent \\
Negative conditioning & ConditioningZeroOut \\
Sampler / scheduler & Euler / simple \\
Sampling steps & 20 \\
CFG / FLUX guidance & 1.0 / 2.5 \\
Reference strategy & Shared reference keyframe \\
Output & Five derivative keyframes \\
\bottomrule
\end{tabular}
\end{table}

\noindent\textbf{First-last-frame Video Generation.}
Wan2.2-I2V-A14B jointly conditions each local generation on its starting keyframe, ending keyframe, and structured dynamic prompt. A high-noise model first produces coarse motion, and a low-noise model refines visual details. Each generated clip contains 81 frames; clips are concatenated within a shot and then composed in storyboard order. Table~\ref{tab:app_video_config} summarizes the settings.

\begin{table}[H]
\centering
\small
\caption{Implementation configuration of First-last-frame Video Generation.}
\label{tab:app_video_config}
\begin{tabular}{ll}
\toprule
Component & Setting \\
\midrule
Video backbone & Wan2.2-I2V-A14B \\
Diffusion stages & High-noise and low-noise models \\
Conditioning & First and last keyframes + dynamic prompt \\
Sampler / scheduler & Euler / simple \\
Sampling steps & 20 \\
CFG scale & 4.0 \\
Output resolution & \(640\times480\) or \(640\times640\) \\
Local clip length & 81 frames \\
Output & Local video clip \(V_k^i\) \\
\bottomrule
\end{tabular}
\end{table}

\section{Detailed Evaluation Protocol}
\label{app:evaluation_details}

Let \(S\) denote an input sketch, \(I_0\) the generated reference frame, and \(\{I_t\}_{t=1}^{T}\) uniformly sampled frames from a generated shot. We sample six frames, including the first and last frames. CLIP image and text encoders are denoted by \(E_{\mathrm{img}}\) and \(E_{\mathrm{text}}\)~\cite{radford2021learning}.

\noindent\textbf{Shot control.}
LPIPS measures perceptual distance between \(S\) and \(I_0\), and CLIP Image Similarity measures their high-level semantic correspondence~\cite{zhang2018unreasonable,radford2021learning}. Edge-F1 measures contour agreement after Canny edge extraction~\cite{canny1986computational}. With binary edge maps \(E_S\) and \(E_{I_0}\), precision and recall are \(P=|E_S\cap E_{I_0}|/|E_{I_0}|\) and \(R=|E_S\cap E_{I_0}|/|E_S|\), and \(\mathrm{EdgeF1}=2PR/(P+R)\).

\medskip
\noindent\textbf{Shot consistency.}
Temporal CLIP averages CLIP image similarity between \(I_0\) and the sampled frames, while Temporal LPIPS averages their perceptual distance. Higher Temporal CLIP and lower Temporal LPIPS indicate stronger intra-shot stability.

\medskip
\noindent\textbf{Story alignment.}
Static Alignment compares the appearance prompt with \(I_0\), while Story Alignment averages the CLIP text-image similarity between the motion prompt and the sampled frames.

\medskip
\noindent\textbf{Local video quality.}
For each predefined event \(e_i\), we find the best CLIP text-image match over the colored reference and sampled frames. A match at or above \(\tau=0.25\) counts as completed. Event Completion averages the binary event outcomes. Dynamic Controllability combines the event success rate \(R_s\), the order rate \(R_o\) of adjacent matched events, and their mean confidence \(R_c\) as \(0.4R_s+0.3R_o+0.3R_c\).
These metrics follow the event realization perspective of StoryEval and the dynamics-aware view of DEVIL~\cite{wang2024storyeval,liao2024devil}.

\medskip
\noindent\textbf{Auxiliary dynamic progression.}
Dynamic Progression measures observable temporal change using the average adjacent-frame CLIP distance, the first-to-last CLIP distance, and the fraction of adjacent pairs above a change threshold. The three terms are combined with weights 0.4, 0.3, and 0.3, respectively. This auxiliary measure is interpreted together with the task-specific control and alignment metrics.

\section{Baseline Implementation Details}
\label{app:baseline_details}

All methods use the full SketchLongVideo evaluation set. For methods accepting one text condition, the input is the concatenation of the appearance prompt and motion prompt. Methods that accept sketches use the same structural input whenever their interface allows it.

\noindent\textbf{SketchVideo.}
The 1KF setting uses the input sketch at its first control frame, while the 2KF setting uses two sparse sketch controls. Both use the released CogVideoX-based inference pipeline~\cite{SK1,yang2024cogvideox}.

\medskip
\noindent\textbf{VidSketch and FlipSketch.}
VidSketch receives one sketch, the concatenated prompt, and the colored reference image required by its interface~\cite{SK2}. FlipSketch receives one static sketch and the same text condition and uses its released inference setup without task-specific adaptation~\cite{SK12}.

\medskip
\noindent\textbf{StoryDiffusion.}
StoryDiffusion receives only the text conditions and generates story keyframes with SDXL~\cite{L9,podell2024sdxl}. The resulting keyframes are animated with the same Wan2.2 backend used by DrawVideo.

\medskip
\noindent\textbf{Wan2.2 variants.}
The text-only variant receives the concatenated text condition without a sketch or intermediate keyframes. The sketch-and-text variant receives the original sketch and text condition directly. Both use the same Wan2.2 configuration as DrawVideo, isolating the contribution of appearance anchoring, action-keyframe expansion, and motion decomposition.

\section{Human Evaluation}
\label{app:human_evaluation}

We conduct a blind and randomized human evaluation with ten participants who have experience in video content, animation, or visual media. For each storyboard, participants see the sketch, appearance prompt, motion prompt, and anonymized outputs from DrawVideo and the seven baselines. They rate each result on a 1--5 Likert scale for structural faithfulness, appearance consistency, motion naturalness, storyboard controllability, and overall quality. Table~\ref{tab:human_study} reports the mean opinion scores. DrawVideo receives the highest mean ratings for structural faithfulness, appearance consistency, storyboard controllability, and overall quality, while remaining competitive in motion naturalness.

\begin{table}[H]
\centering
\scriptsize
\setlength{\tabcolsep}{3.5pt}
\caption{Human evaluation using 1--5 Likert scores. Best results are shown in bold.}
\label{tab:human_study}
\begin{adjustbox}{width=\linewidth}
\begin{tabular}{lccccc}
\toprule
Method & Structural & Appearance & Motion & Control & Overall \\
\midrule
SketchVideo (1KF) & 2.37 & 2.51 & 2.41 & 2.35 & 2.41 \\
SketchVideo (2KF) & 2.45 & 2.97 & 2.81 & 2.76 & 2.75 \\
VidSketch & 3.42 & 2.38 & 2.24 & 3.03 & 2.77 \\
FlipSketch & 2.08 & 1.98 & 2.13 & 2.58 & 2.19 \\
StoryDiffusion & 2.05 & 2.48 & 2.86 & 2.58 & 2.49 \\
Wan2.2 (text only) & 2.22 & 2.86 & 3.42 & 2.45 & 2.74 \\
Wan2.2 (sketch+text) & 3.42 & 3.31 & \textbf{3.63} & 3.03 & 3.35 \\
\midrule
DrawVideo & \textbf{3.69} & \textbf{3.72} & 3.56 & \textbf{3.79} & \textbf{3.69} \\
\bottomrule
\end{tabular}
\end{adjustbox}
\end{table}

\section{Additional Qualitative Results}
\label{app:additional_results}

\begin{figure}[H]
  \centering
  \includegraphics[width=\linewidth]{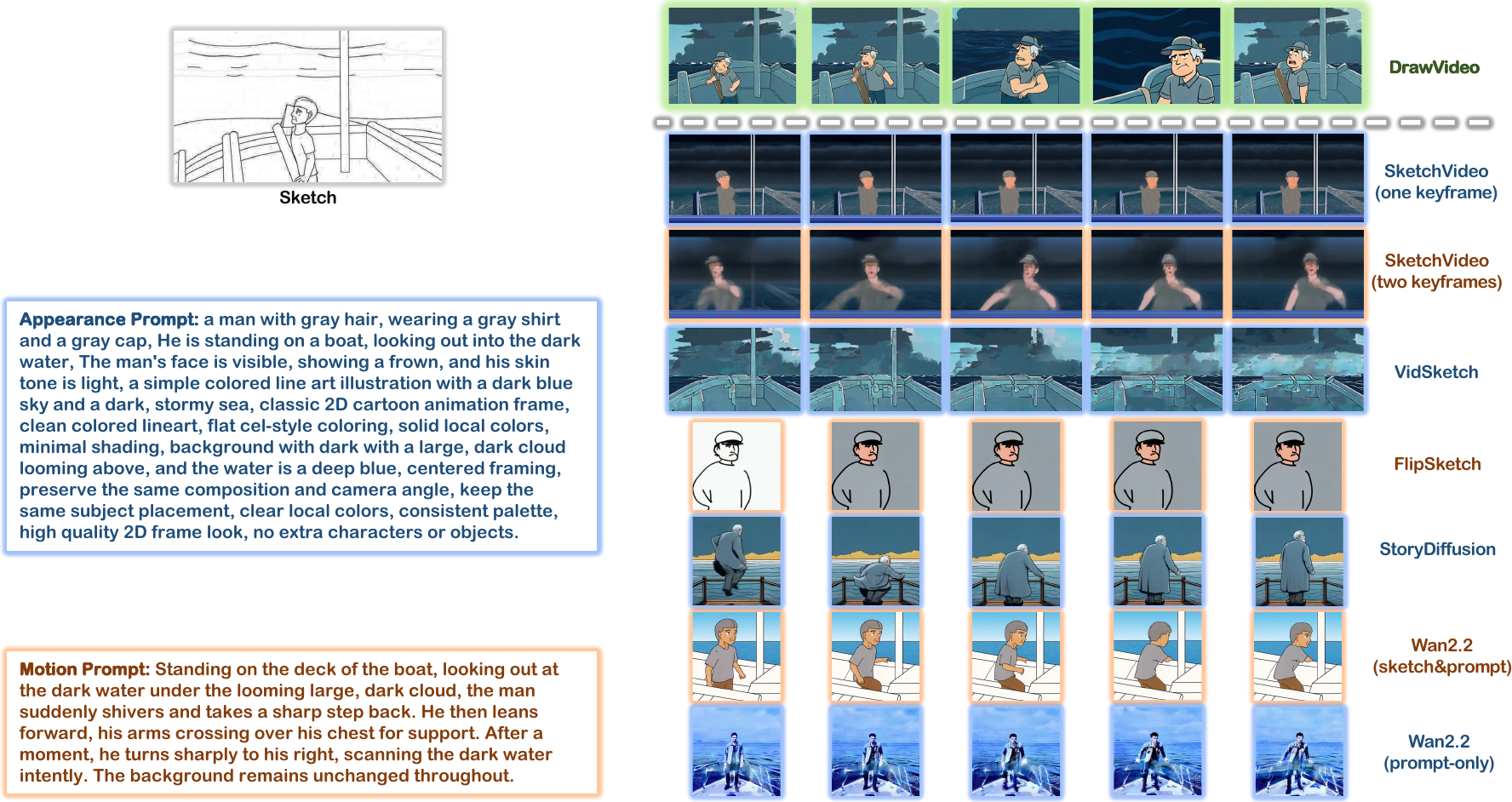}
  \caption{Qualitative comparison on a storyboard set on a boat.}
  \label{fig:app_comparison_boat}
\end{figure}

\begin{figure}[H]
  \centering
  \includegraphics[width=\linewidth]{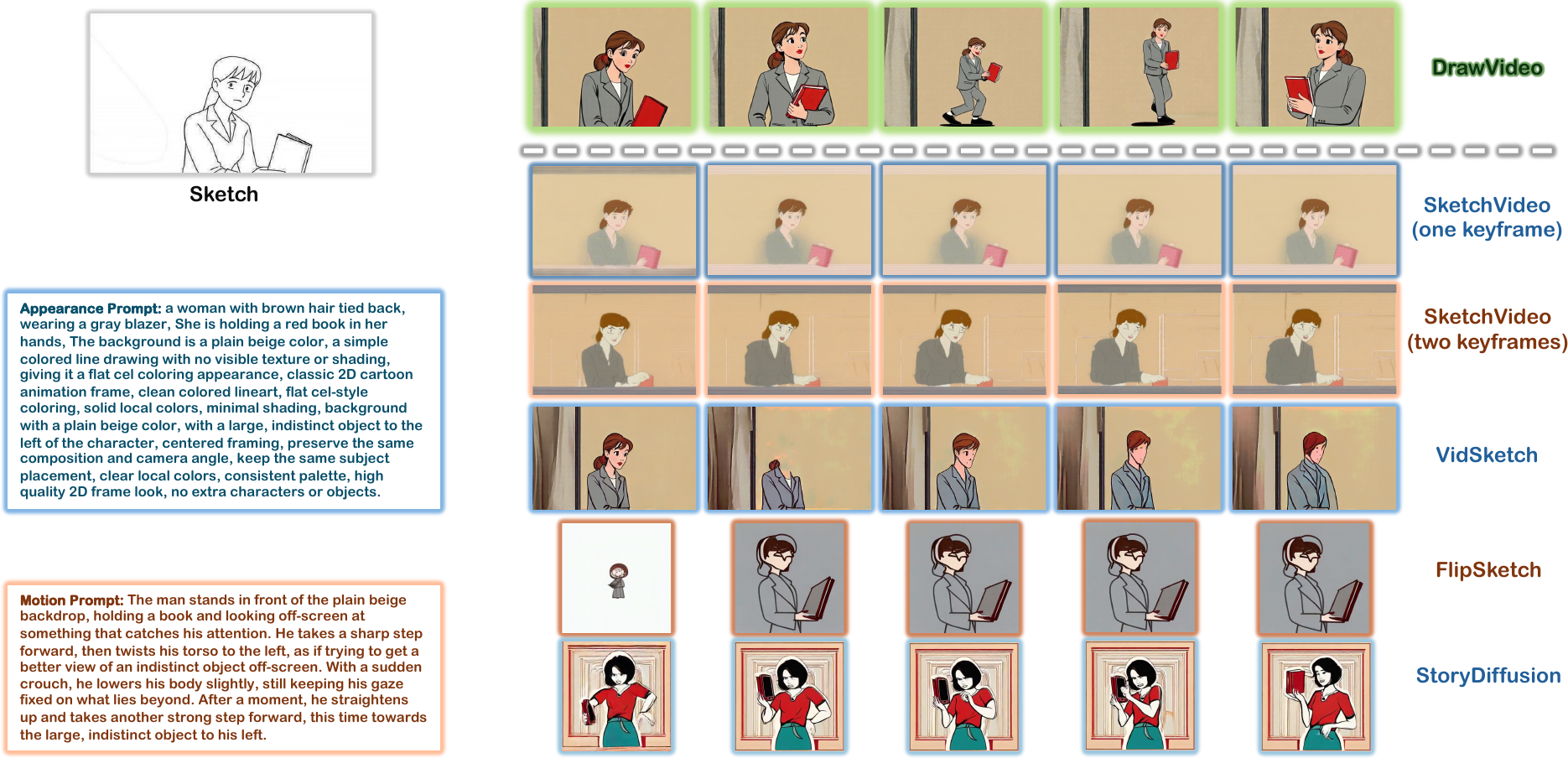}
  \caption{Additional qualitative comparison 1.}
  \label{fig:app_additional_1}
\end{figure}

\begin{figure}[H]
  \centering
  \includegraphics[width=\linewidth]{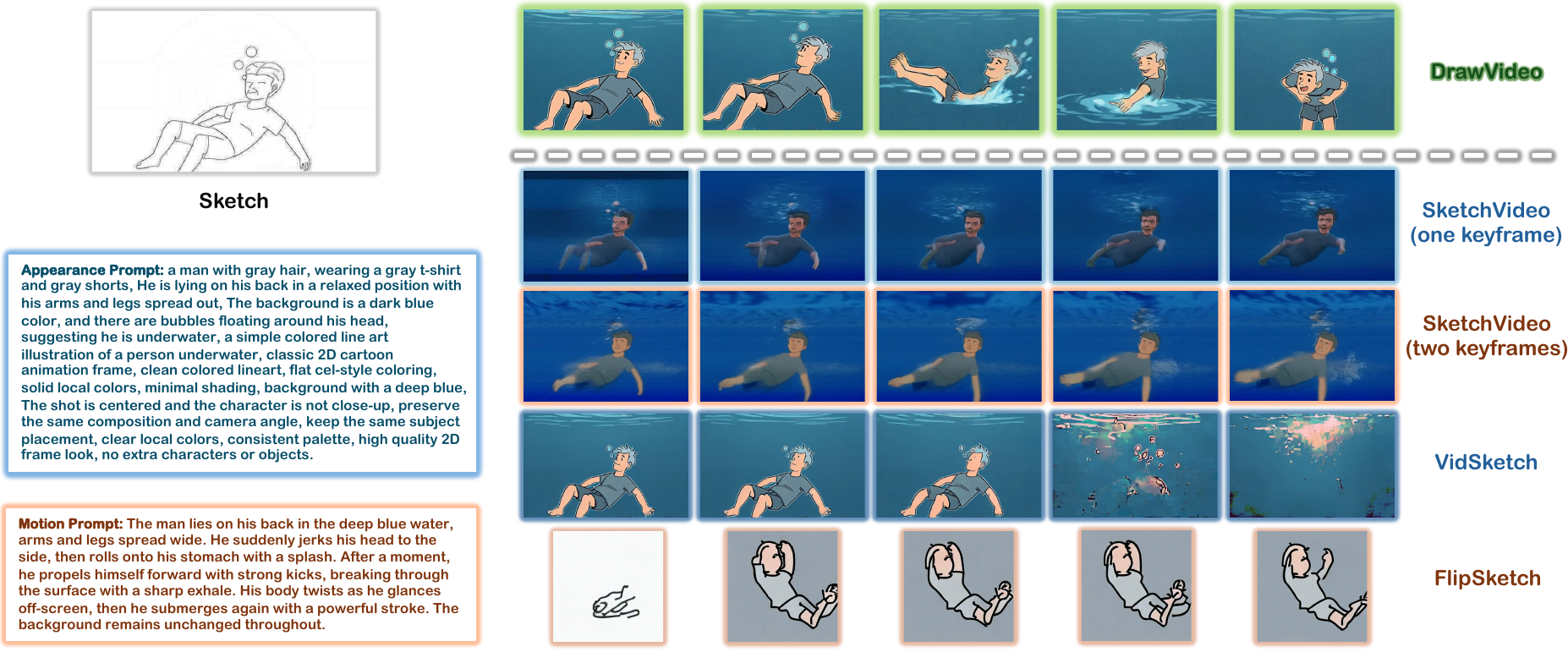}
  \caption{Additional qualitative comparison 2.}
  \label{fig:app_additional_2}
\end{figure}

\begin{figure}[H]
  \centering
  \includegraphics[width=\linewidth]{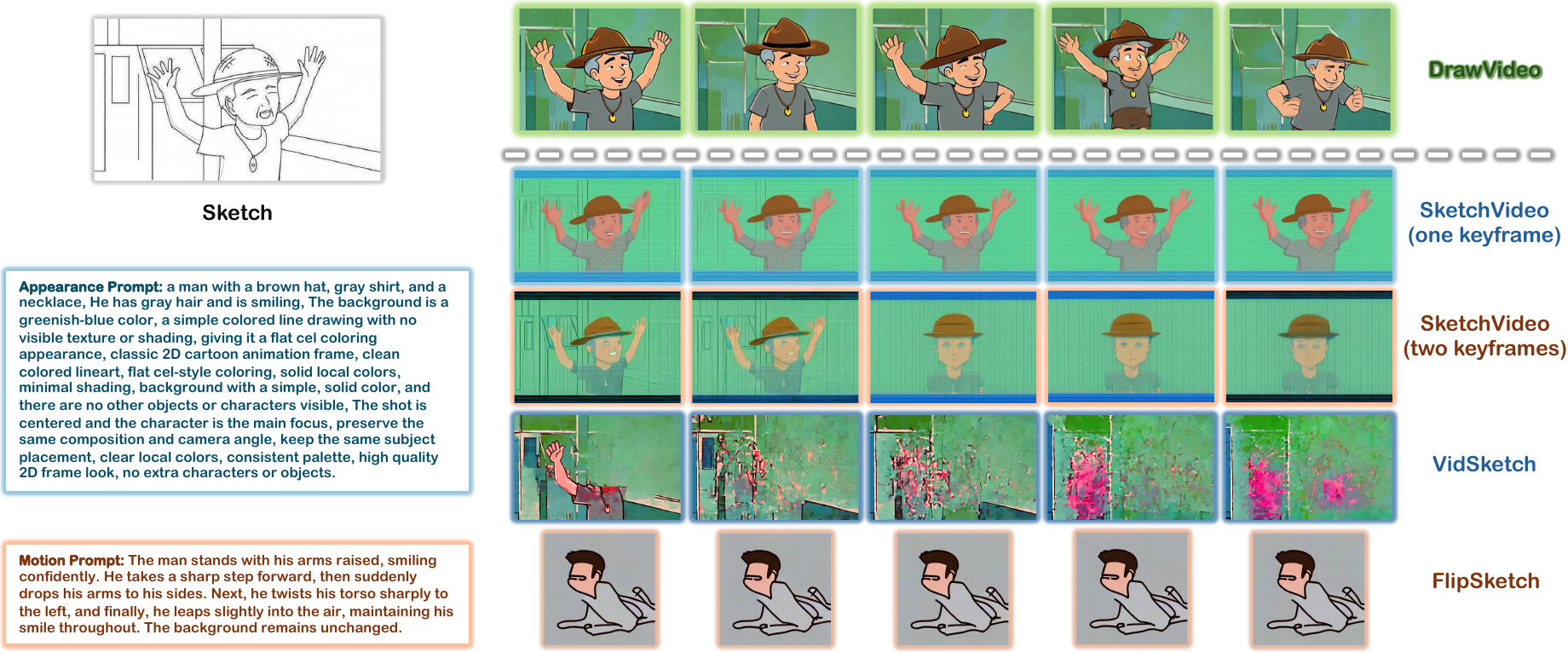}
  \caption{Additional qualitative comparison 3.}
  \label{fig:app_additional_3}
\end{figure}

\begin{figure}[H]
  \centering
  \includegraphics[width=\linewidth]{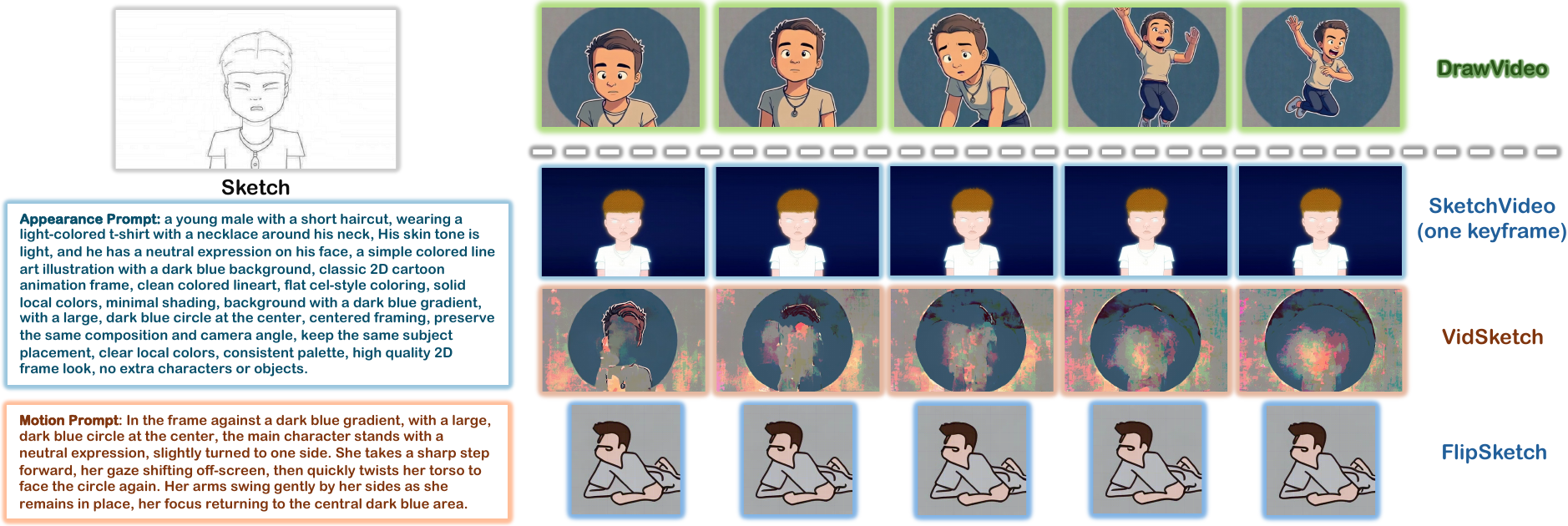}
  \caption{Additional qualitative comparison 4.}
  \label{fig:app_additional_4}
\end{figure}

\begin{figure}[H]
  \centering
  \includegraphics[width=\linewidth]{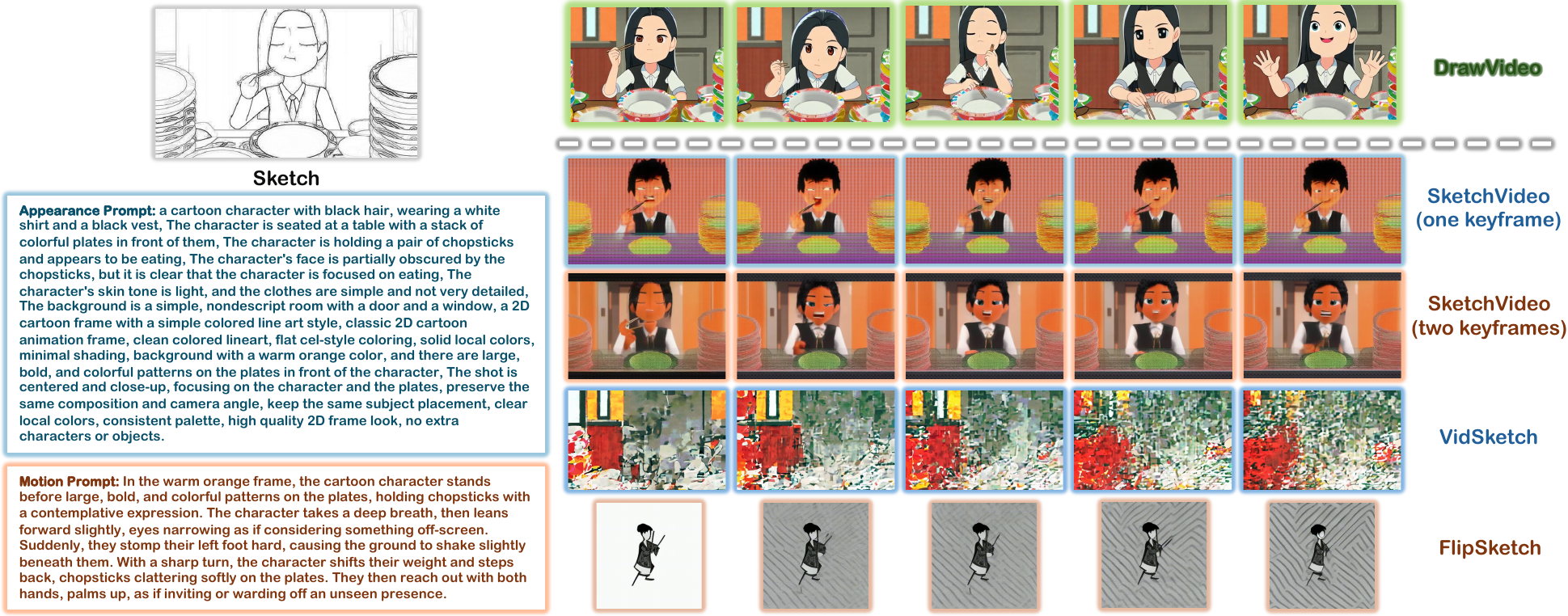}
  \caption{Additional qualitative comparison 5.}
  \label{fig:app_additional_5}
\end{figure}

\begin{figure}[H]
  \centering
  \includegraphics[width=\linewidth]{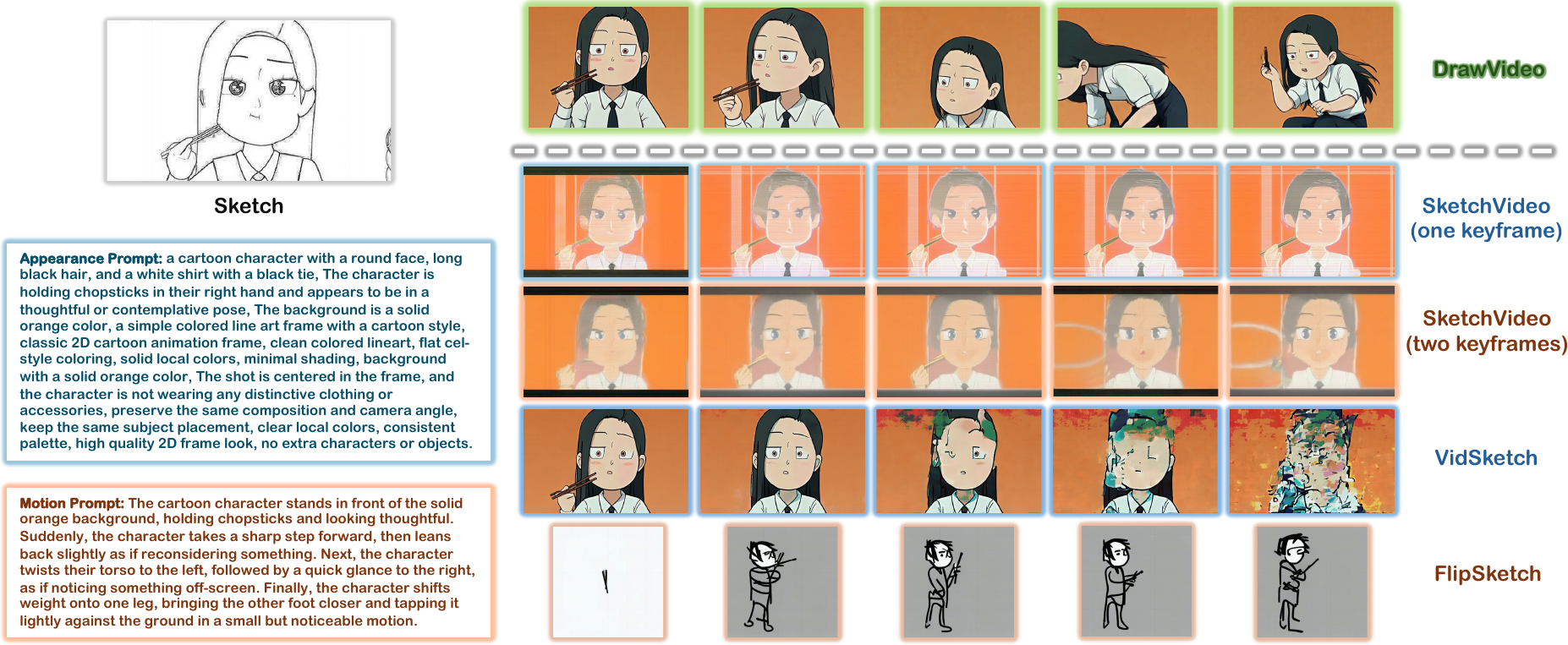}
  \caption{Additional qualitative comparison 6.}
  \label{fig:app_additional_6}
\end{figure}

\begin{figure}[H]
  \centering
  \includegraphics[width=\linewidth]{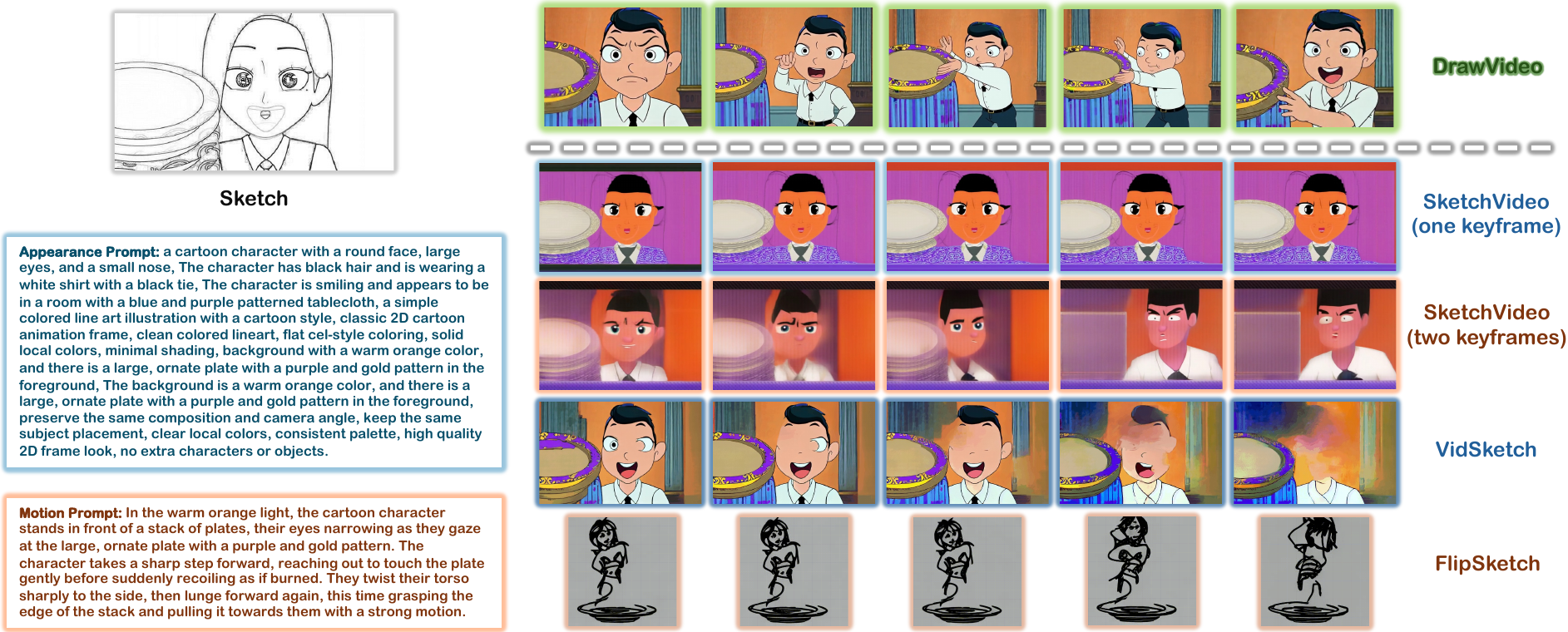}
  \caption{Additional qualitative comparison 7.}
  \label{fig:app_additional_7}
\end{figure}

\begin{figure}[H]
  \centering
  \includegraphics[width=\linewidth]{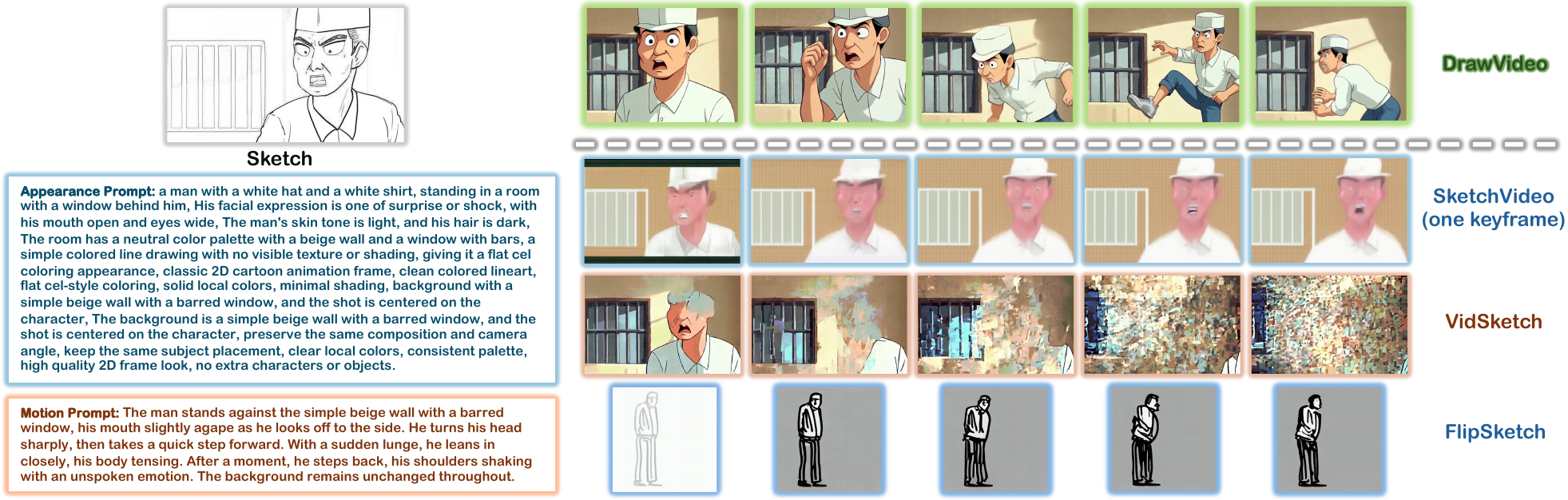}
  \caption{Additional qualitative comparison 8.}
  \label{fig:app_additional_8}
\end{figure}

Figure~\ref{fig:app_comparison_boat} provides an additional comparison on a boat scene, complementing the phone-interaction and fine-grained eating comparisons in Figures~\ref{fig:comparison} and \ref{fig:comparison_eating}. Figures~\ref{fig:app_additional_1}--\ref{fig:app_additional_8} provide eight further comparisons across different storyboard conditions. All figures use the same baseline ordering and evaluation setting as the main paper.

\end{document}